\documentclass[aps,prl,twocolumn]{revtex4-1}

\usepackage{dcolumn}
\usepackage{graphicx}

\begin{document}
\title{Statistics of the General Circulation from Cumulant Expansions}
\author{J .B. Marston}
\affiliation{Department of Physics, Brown University,
Providence, Rhode Island 02912-1843 U.S.A.}
\maketitle

 \vspace*{-0.8cm}

Climate varies on decadal and longer time
scales, yet climate simulations advance at intervals measured in
minutes or even seconds. In 1967, however, Edward Lorenz observed: 
``More than any other
theoretical procedure, numerical integration is also subject to the
criticism that it yields little insight into the problem. The computed
numbers are not only processed like data but they look like data, and
a study of them may be no more enlightening than a study of real
meteorological observations. An alternative procedure which does not
suffer this disadvantage consists of deriving a new system of
equations whose unknowns are the statistics
themselves.''\cite{Lorenz67}   We show here that such direct statistical simulation
(DSS) is able to reproduce statistics obtained by the traditional
route of time averaging direct numerical simulations (DNS).  

\begin{figure}[t]
\includegraphics[width=3.0in]{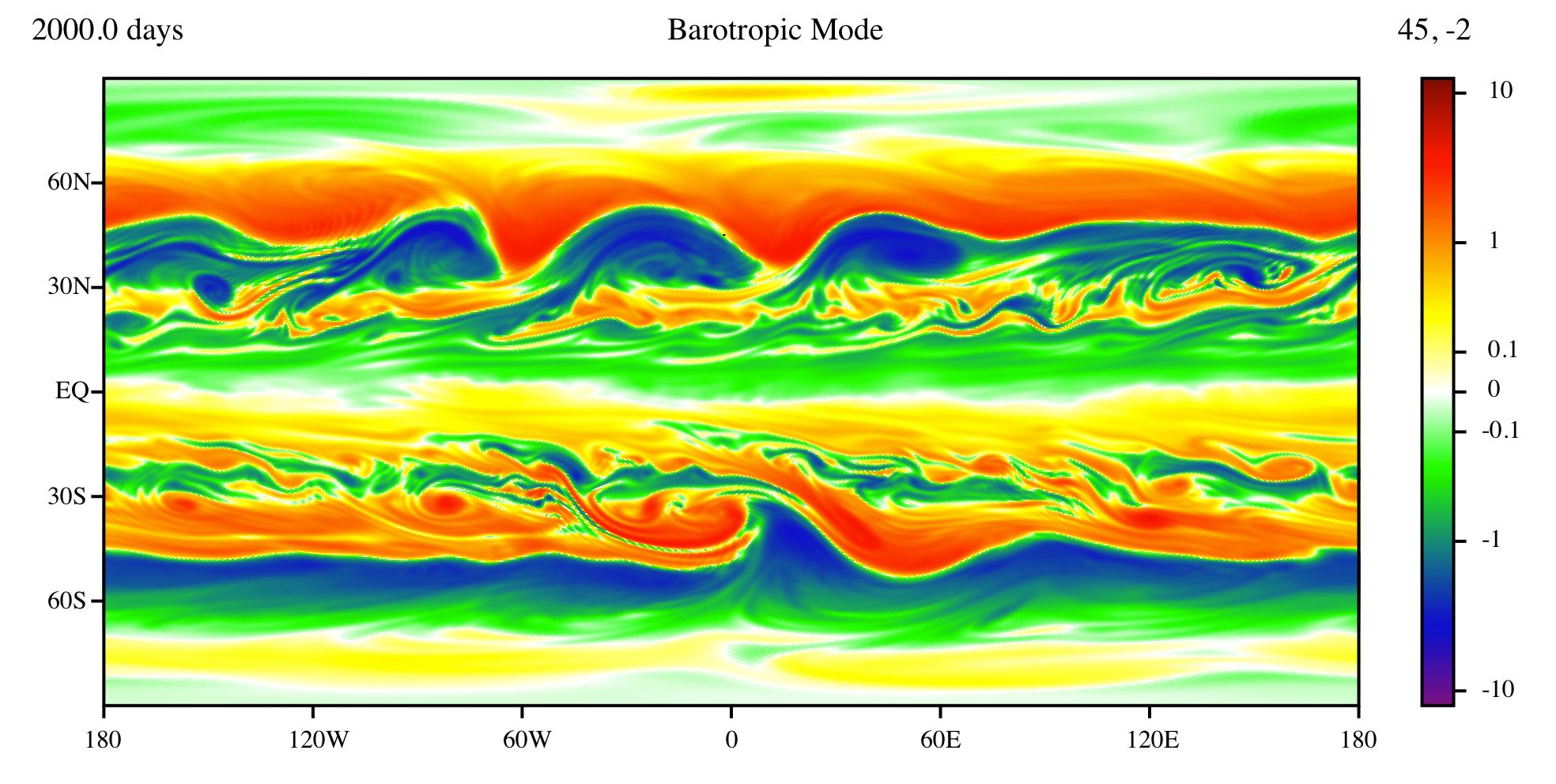}
\caption{\label{figure1} Instantaneous relative vorticity field of the two-layer model on a spherical geodesic grid with 163,842 cells.}
\end{figure}

DSS of damped and driven systems may be formulated by a Reynolds decomposition of the dynamical variables $q_i$ into the sum of a mean value and a fluctuation (or eddy):
$q_i = \langle q_i \rangle + q_i^\prime\  \ {\rm with}~ \langle q_i^\prime \rangle = 0$. 
The first two equal-time cumulants are then $c_i \equiv \langle q_i \rangle$ and 
$c_{ij} \equiv  \langle q_i^\prime q_j^\prime \rangle = \langle q_i q_j \rangle - \langle q_i \rangle \langle q_j \rangle$.
The equation of motion for the cumulants form an infinite hierarchy that must be truncated.
The simplest closure, CE2, is to set the third and higher cumulants equal to
zero, amounting to the neglect of eddy-eddy
interactions. Here we extend the work of Ref. \cite{Marston:2008p1} to a two-layer quasigeostrophic model originally formulated on the $\beta$-plane\cite{Lee:1996p193} but generalized to the rotating sphere.    
DNS is carried out in real space on a spherical geodesic grid.\cite{Heikes:1995p113}   A typical instantaneous flow is shown in Fig. \ref{figure1}.  

\begin{figure}[t]
\includegraphics[width=3.0in]{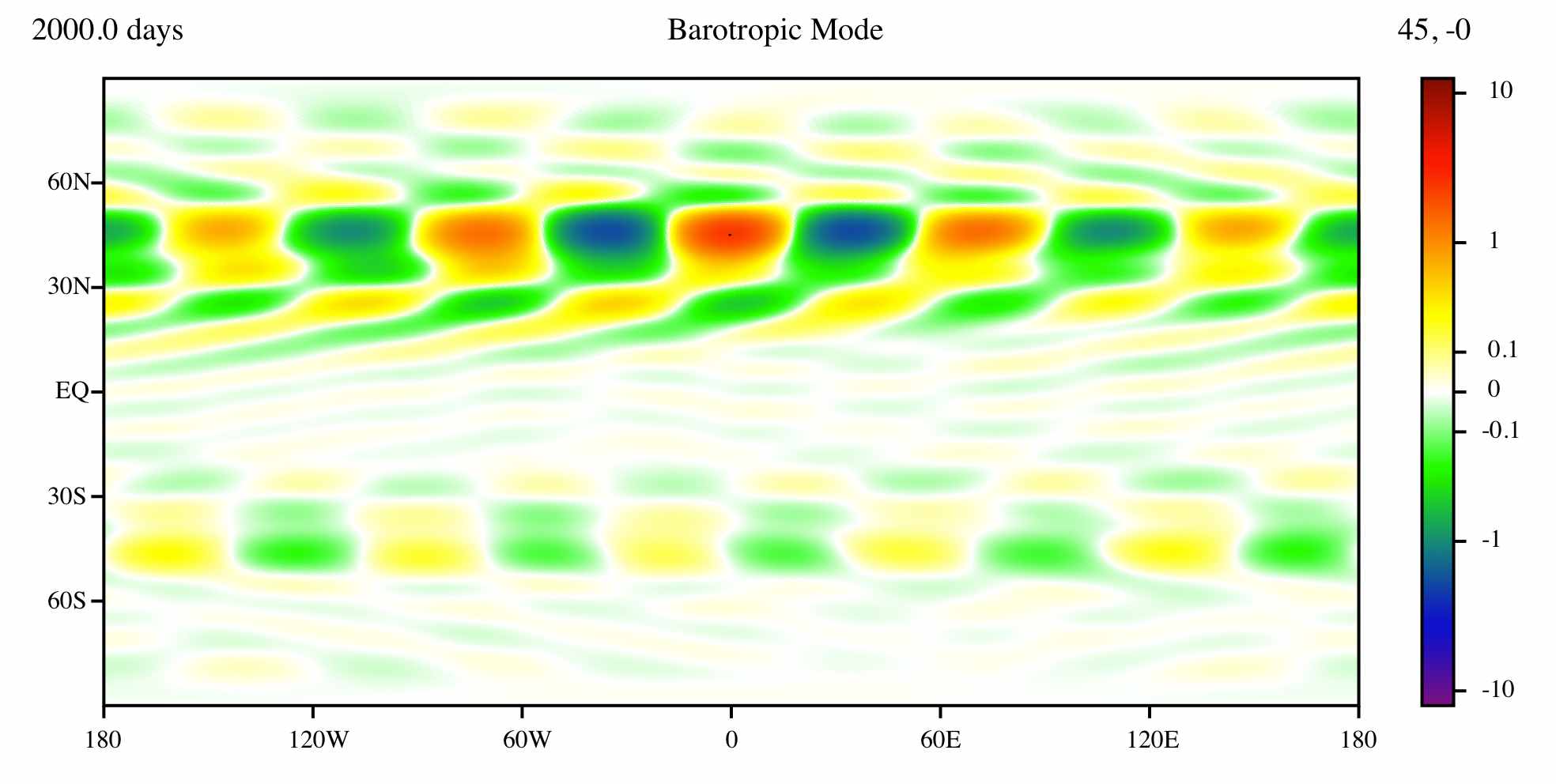}
\includegraphics[width=3.0in]{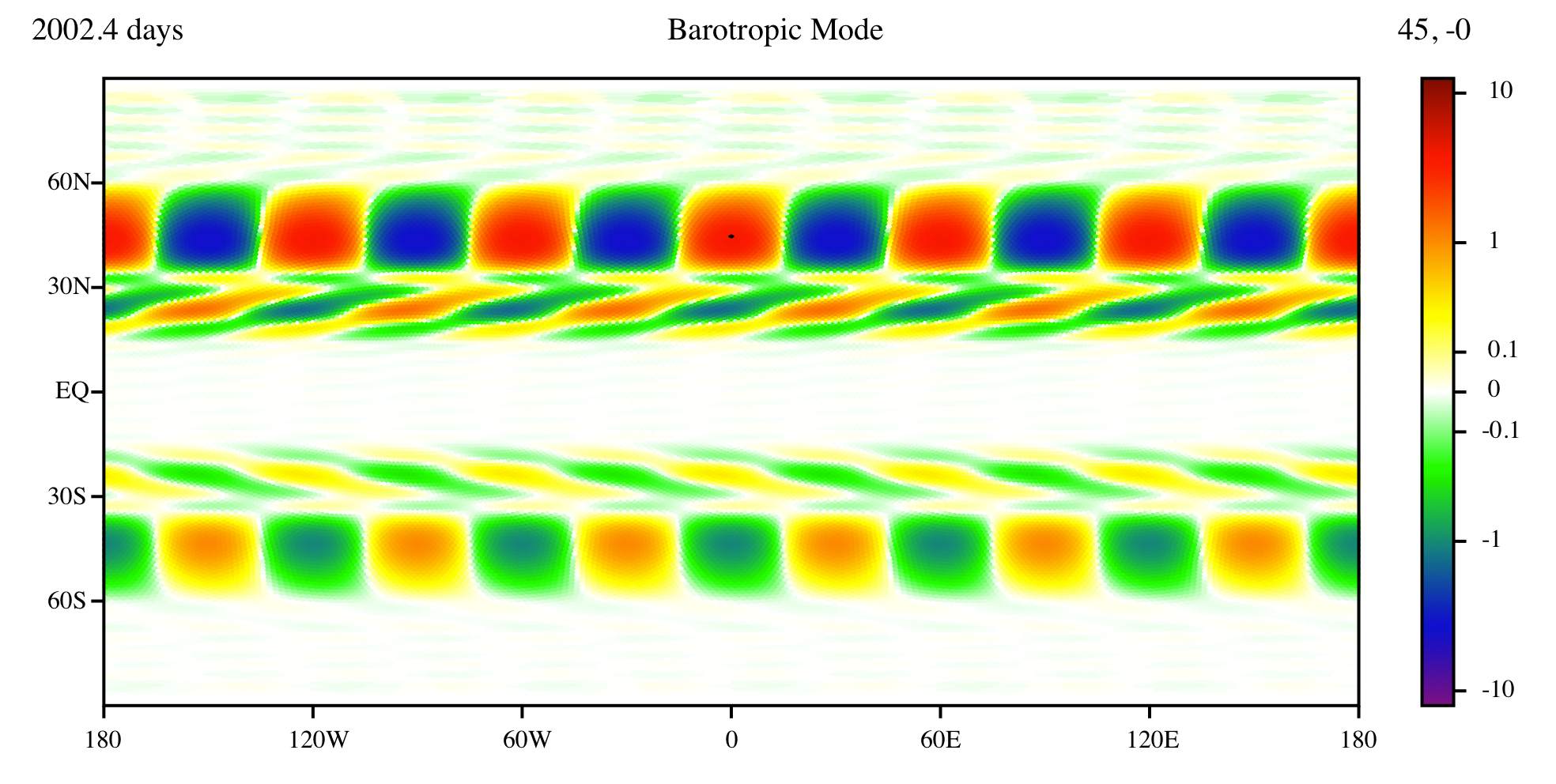}
\caption{\label{figure2} The two-point second cumulant of the relative vorticity field with respect to a reference field point at latitude $45^\circ$ and along the prime meridian. Top:  DNS.  Bottom: CE2.}
\end{figure}

\begin{figure}[t]
\includegraphics[width=3in]{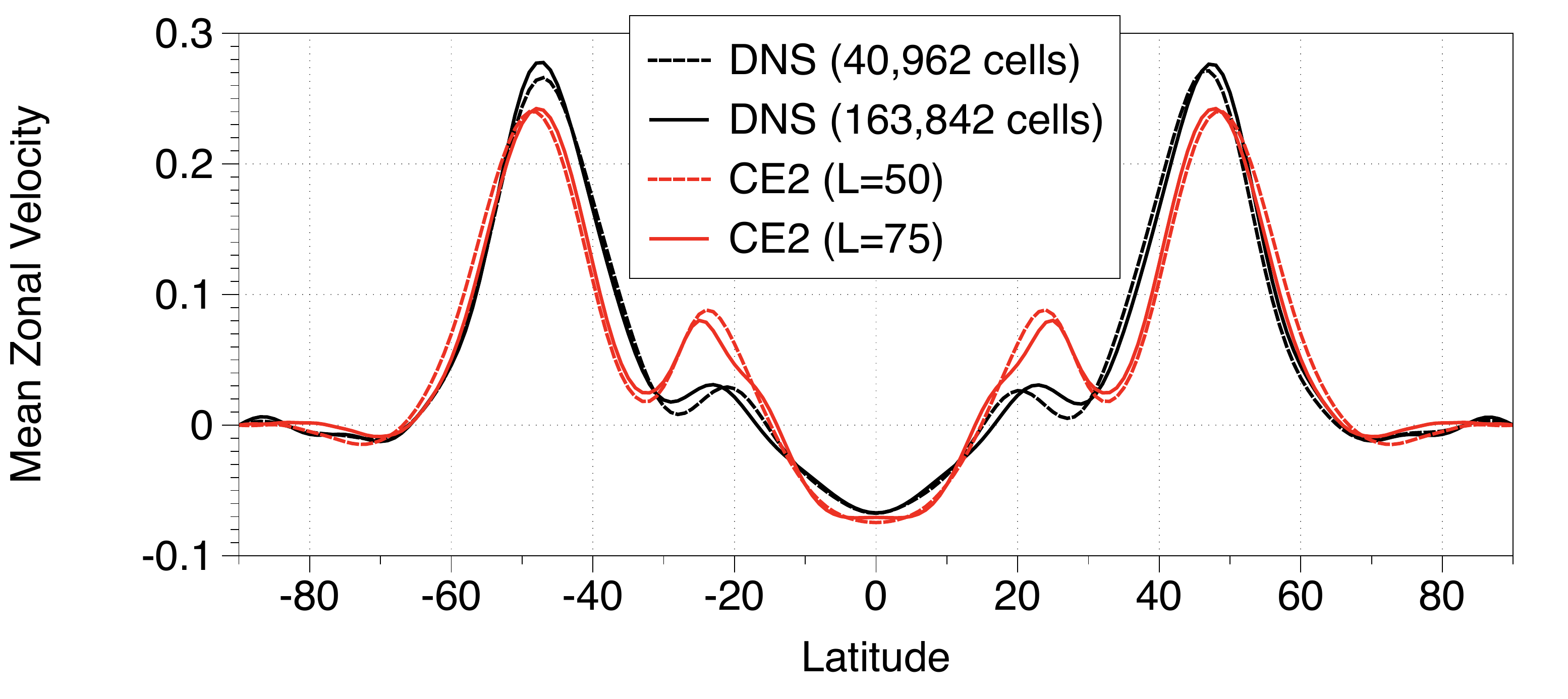}
\caption{\label{figure3} Mean zonal velocity, averaged over both layers, as calculated by DNS and CE2 at two different resolutions.}
\end{figure}

Fig. \ref{figure2}
compares the two-point vorticity correlation function as calculated by DNS and CE2.  Storm tracks are immediately apparent from the strong mid-latitude
correlations and the jets fluctuate mainly at zonal wavenumber 5 and 6.   CE2
captures long-range correlations (``teleconnection patterns'') and
inter-hemisphere correlations are weak, as expected.
A quantitative comparison of the zonal velocities, as calculated by DNS and CE2, is presented in Fig. \ref{figure3} where the midlatitude westerlies and the tropical trade winds are evident. 
The comparison suggests that the large-scale flows are not so nonlinear as to preclude their description by the cumulant expansion.  
Stratification and shearing act together to weaken the nonlinearities in large-scale flows.\cite{OGorman:2007p76}  It is these features of the general circulation that distinguishes it from the highly nonlinear problem of 3D isotropic and homogenous turbulence, and makes DSS possible.    

I thank P. Kushner, T. Schneider and S. Tobias.  This work was supported in part by NSF grant Nos. DMR-0605619 and PHY-0551164.

\end{document}